# On the Capacity of the Peak Limited and Band Limited Channel

Michael Peleg [1] and Shlomo Shamai [1]

1    Technion - Israel Institute of Technology, Department of Electrical and Computer Engineering

**Abstract:** We investigate the Peak-Power Limited (PPL) Additive White Gaussian Noise (AWGN) channels in which the signal is band-limited, and its instantaneous power cannot exceed the power $P$. This model is relevant to many communication systems; however, its capacity is still unknown.  We use a new geometry-based approach which evaluates the maximal entropy of the transmitted signal by assessing the volume of the body, in the space of Nyquist-rate samples, comprising all the points the transmitted signal can reach. This leads to lower bounds on capacity which are tight at high Signal to Noise Ratios (SNR). We found lower bounds on capacity, expressed as power efficiency, higher than the known ones by a factor of 3.35 and 8.6 in the low pass and the band pass cases respectively. The gap to the upper bounds is reduced to a power ratio of 1.5. The new bounds are numerically evaluated for FDMA-style signals with limited duration and also derived in the general case as a conjecture. The penalty in power efficiency due to the peak power constraint is roughly 6 dB at high SNR. Further research is needed to develop effective modulation and coding for this channel.

**Keywords:** peak power; capacity; AWGN; band limited; entropy;

## 1   Introduction

We investigate the Peak-Power Limited (PPL) Additive White Gaussian Noise (AWGN) channels in which the signal is band-limited, and its instantaneous power cannot exceed the power $P$. This model is relevant to many systems in which the peak power is limited by the power amplifier at the transmitter. The model became even more important with the introduction of Digital Pre-Distortion (DPD), e.g. [1] and [2], which linearizes the power amplifier up to its maximal transmit power, thus causing it to perform as an ideal soft limiter. Clearly, the capacity limits of this channel are of major practical interest, e.g. the optimization in section 3.6 of [2], and a discussion in [3]. Indeed, Shannon analysed this channel and presented lower and upper bounds on capacity already in [4]. With the exact capacity of the classical Average Power Limited (APL) channel found by Shannon [4] and used widely for tens of years, the PPL channel capacity was studied only sparsely yielding lower and upper bounds on capacity with a wide gap in between, see [5],[6] and Table 1. We think the reason is the difficulty to analyse this channel as suggested already in [4]. The importance of limiting the peak power is reflected also in many works analysing and reducing the Peak to Average Power Ratio (PAPR), e.g. [7], [8] and [9]. The impact of the peak power limit is classical in many communications settings, since the beginning of the wireless communication era, and it is relevant to a variety of practical communications models, as for example fading channels and the like [3].The problem investigated here is related to communication over the Constrained Gaussian Channel (CGC) [10], [11][12] in which a wideband peak limited signal is fed into a transmit filter in the transmitter. We show below that the capacity of the CGC is an upper bound of the capacity of the PPL channel. The review [13] presents and categorizes a wide range of modulation schemes with different types of peak limit including the CGC and the PPL models.

There are two known types of upper bounds on the capacity of the PPL channel. The first one uses the result of [14] on the Power Spectral Density (PSD) of unit processes, which are the inputs to the CGC channel, to derive upper bound on capacity [12] of the CGC channel which are also valid for the PPL channel. In [15] the approach is specified to the PPL channel gaining additional insights. The second type of upper bounds releases the constraint on the peak power by applying it only to samples of the signal taken at the Nyquist sampling rate and then computing capacity based on the Nyquist rate samples being sufficient statistics of the received signals. We denote this approach here as the "sampled discrete analysis". This is introduced in [4] and used in [5] utilizing the capacity of the scalar peak limited channel derived in [16]. The known lower bounds on capacity are obtained by achievability schemes based on identically and independently distributed ( i.i.d.) symbols with optimized pulse shapes, see [4],[5] and [6].

In this work we provide numerical evaluation of a lower bound on capacity which is valid for Cyclic Prefix assisted Frequency Domain Equalization (CP-FDE) signaling of length of up to 100 channel symbols. The CP-FDE signals are not strictly band-limited because they are limited in time, however, they are practically band limited in the sense of zero inter-channel interference between users if the rules for cyclic prefix are adhered to, thus enabling spacing adjacent users to channels with no frequency gaps in between. This is applied for example in the multiuser uplink of the Long-Term Evolution (LTE) mobile communications system using the Single-Carrier FDMA (SC-FDMA) [17]. Furthermore, we present a lower bound on the capacity of the general PPL channel which is a conjecture due to two analytical approximations. We provide lower bounds on the capacity improved about 5 dB and more relative to [5] and [6]. We show that our lower bounds are tight at asymptotically high Signal to Noise Ratio (SNR) while it is well known, [4], that at very low SNR the PPL capacity approaches the APL one. We investigate both the real-valued channels which model low-pass signals and the complex-valued channels modelling band-pass signals.

The lower bounds in [4],[5] and [6] were obtained using i.i.d. symbols. Our new approach utilizes dependencies between symbols to increase capacity while meeting the peak power constraint. Modern efforts at PAPR reduction use diverse methods, frequently adapting the transmission per each individual information sequence resembling coding, e.g. [9]. We found that the signals emerging in our new bounds utilize only a small subset of possible symbol sequences selected by the peak power constraint resembling in a way coding in which only a small subset of all possible binary sequences selected by the parity check matrix are valid codewords.

Our new approach is geometry-based, it evaluates the maximal entropy of the transmitted signal by assessing the volume of the body comprising all the points the transmitted signal can reach in the space of Nyquist-rate samples. This is related to the technique introduced in [18] over the CGC channel.

Notation: Log is the natural logarithm unless stated otherwise. Differential entropy is denoted by $h$, $E$ denotes the statistical expectation. The $N$-dimensional vector space of real variables is denoted $R^N$. Probability Density Function (PDF) of $x$ is $p_x(x)$ or $p(x)$.

## 2 System model

We begin with the real-valued channel. The system is presented in Figure 1. The encoder produces a real-valued low-pass signal $x(t)$ in the frequency band $|f|<B$. The signal is peak-limited, that is, $|x(t)| \leq \sqrt{P}$ for all $t$. The signal passes an AWGN channel and is decoded. The channel output $y$ is

$$y(t) = x(t) + n(t) \tag{1}$$

where $n(t)$ is a white Gaussian noise with power spectral density $N_0$ ($0.5N_0$ two-sided) and power $\sigma_n^2 = N_0B$. The Nyquist interval is $T=0.5/B$. The signal to noise ratio is defined as $\rho = \frac{P}{BN_0}$ . We seek bounds on the capacity which is the maximal Mutual Information (MUI) denoted $I(x;y)$ per Nyquist interval. The Nyquist-rate sampled $x(t)$ is denoted by the vector $\mathbf{x} = (x_1 \ldots x_n \ldots x_N)$ of length $N$.

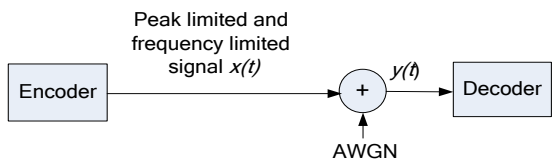

**Figure 1.** System model of the PPL channel.

The capacity in bits per Nyquist interval of a similar APL system in which the peak power limit $P$ is replaced by the average power limit is the famous [4]

$$C_a = 0.5 \log_2 \left(\frac{P}{N_0B} + 1\right). \tag{2}$$

As stated in the introduction, the capacity of the CGC channel is an upper bound of the capacity of the PPL channel, the proof is by Lemma 1 of [10] which implies that any signal permitted at the channel input by the PPL model is valid also under the CGC model.

# 3 Analysis

## 3.1 General analysis

Lower bound on capacity can be obtained from the differential entropy $h(x)$ of the transmitted signal $\mathbf{x}$ via the Entropy Power Inequality (EPI), e.g.[19], as done e.g. in [10]. The derivation is presented in Appendix A for completeness. The lower bound $\gamma$ on the power loss ratio of the PPL channel relative to that of APL channel (2) valid at all SNRs is defined in the sense of ( 3)

$$C \geq 0.5 \log_2\left(\frac{\gamma \cdot P}{N_0 B} + 1\right).$$ ( 3)

It is shown in Appendix A that the following holds:

$$\gamma = \frac{P^e}{P} \leq 1$$ ( 4)

where $P^e$ denotes the entropy power of $x$ defined as

( 5)

$$P^e = \frac{1}{2\pi e} \cdot e^{\frac{2}{N}h(x)}$$

The ratio $\gamma$ is pre-SNR factor in [6] and it is unity for the APL channel, leading from ( 3) to (2). If the transmitted vector of the Nyquist-rate samples $\mathbf{x}$ is confined to some region of $R^N$ with a volume $V_x$ then the maximal $h(x)$ is the logarithm of the volume $V_x$ and is achieved by uniform distribution of $\mathbf{x}$ over $V_x$. As shown in the appendix, under the uniform distribution we have

$$P^e = \frac{1}{2\pi e} V_x^{\frac{2}{N}}$$ ( 6)

Combining with
( 4) yields

$$\gamma = \frac{V_x^{\frac{2}{N}}}{2\pi e}$$ ( 7)

evaluated for signals with peak power $P$=1, see Appendix A.

To provide an upper bound on $\gamma$, the peak power limit can be applied on the Nyquist rate samples only and not on the signal in between as done in [5] yielding $\gamma = \frac{2}{\pi e}$, see Appendix A for further explanation.

The lower bound ( 3), ( 6) and ( 7) which is valid for all SNRs, was shown by [16] to be tight at asymptotically high SNR for the one-dimensional system analysed in [16] ; it is tight at asymptotically high SNR in our case too as shown below equation (13).

We follow the method of evaluating the volume $V_x$ presented in [18]. As in [18], all the peak limited signals form a convex body where convex means that for any pair $\mathbf{x}_1$, $\mathbf{x}_2$ in $V_x$, any linear combination of the two vectors $a\mathbf{x}_1+b\mathbf{x}_2$, with $a$, $b$ positive and $a_1+a_2$=1, is in $V_x$. This holds in our case since the absolute value of the linear combination is upper bounded as

$$|ax_1(t) + bx_2(t)| \leq a|x_1(t)| + b|x_2(t)|; \quad a + b = 1, a, b > 0.$$

We seek the volume $V_x$ of the $N$-dimensional convex set $\mathbf{x}$ which includes the origin. Denote by $r$ the Distance From the Origin (DFO) to the set surface along the direction of some vector $\mathbf{x}$. We denote $r$ as DFO while in [20] the term radial function is used. Denote the PDF of $r$ by $p(r)$ for angles $\theta$ from the origin to the surface selected randomly and uniformly over the $N$-1-sphere. Then the volume is

$$V_x = E_\theta(V_N^u \cdot r^N)$$ ( 8)

with $E$ denoting expectation and $V_N^u$ denoting the volume of the $N$-dimensional ball with unit radius. The volume equation is equivalent to proposition 1.13 in [20] where $r$ is denoted radial function. In [20], the expectation is replaced by an integral over all directions, this is equivalent since in our calculation the directions are uniformly distributed over the unit $N$-1-sphere. The conversion between ( $8$) here and proposition 1.13 in [20] utilizes also the relation $S_{N-1}^u = N \cdot V_N^u$ where $S_{N-1}^u$ is the area of the unit $N$-1 sphere. The expression in [20] is stated and valid for star-shaped bodies which are a generalization of the convex bodies which include the origin used in this work.

To estimate the volume, we draw random vectors $\mathbf{x}$ with angles $\theta$ spread uniformly over the sphere, calculate $r$ for each, substitute it into ( $8$) and compute the average. The random vectors are generated by a method taken from [21], which is vectors of independent random Gaussian components. The estimation of $V_x$ is illustrated in the following figure in two dimensions.

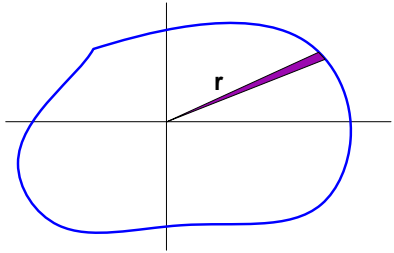

**Figure 2.** Illustrating the estimation of volume of a convex set.

This approach is applied to our problem by:

1.  Generating, numerically or analytically, a signal at a random direction distributed uniformly on the $N$-1 sphere. This is done by drawing all the samples independently from a Gaussian distribution.
2.  Scaling the generated signal to observe the peak limit of $x(t)$ at any time and sampling it at Nyquist rate to obtain the vector $\mathbf{x}$. The resulting length $L$ of $\mathbf{x}$ determines the local radius $r$ (DFO) in this direction.
3.  Averaging as defined in ( $8$) over many directions yields the volume of the convex body.

The analysis needs to be careful because, in most cases, the peak of the signal before normalization will be very large (maximum of many Gaussian variables) and the volume will be determined mostly by the minority of signals in which the peak is moderate. This minority is vanishing with growing $N$. To address this issue, we shall perform the analysis on blocks of length of $NT$ and then estimate the limit as $N$ approaches infinity.

## 3.2 Sampled discrete analysis

The analysis of maxima of continuous Gaussian processes is difficult and will require approximations, [22]. For an initial analysis, we shall peak-limit the signal using only the $N$ Nyquist rate samples. This is the same as the problem of a sampled system using discrete power-limited symbols treated in [4],[5] and[6] and serves us to develop our analysis method. The analysis is generalized to continuous signals further below. We need the PDF of

$$z = \max_{i \ni \{1...N\}} |x_i|$$

This is available from the theory of order statistic e.g. [23], [24]. From [23] eq. 2.1.6 we have

$$p_z(z) = N \cdot F_{|x|}(z)^{N-1} \cdot p_{|x|}(z)$$

where $p$ and $F$ denote the PDF and Cumulative Distribution Function (CDF) respectively. For Gaussian $x$ with

$$F_X(x) = P(x \le X) = 1 - Q(x)$$

this yields after few standard steps:

$$P_Z(z) = N[1 - 2Q(z)]^{N-1} \frac{2}{\sqrt{2\pi}} e^{-\frac{z^2}{2}} \qquad (9)$$

The distribution $p(z)$ is presented for various $N$ and verified by simulation in Figure 3.

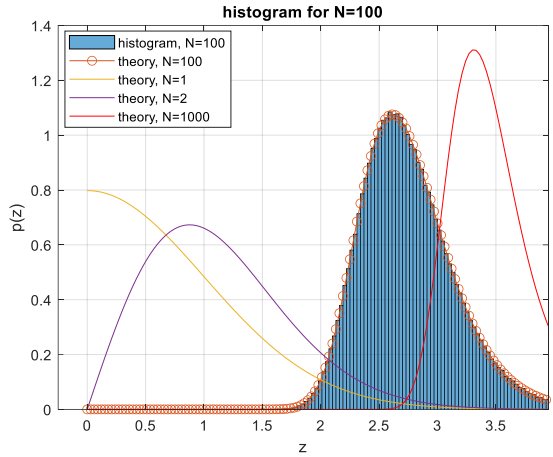

**Figure 3.** The PDF of the maxima $z$ of absolute values of $N$ Gaussian random variables $x_i$.

To evaluate the DOF $r$, we need the PDF of the vector $\mathbf{x}$ of the Gaussian variables when the maxima $z$ of the absolute value of its elements is given. This is related to order statistics e.g. [23], [24], however such approach leads to too complicated analysis. We use the following approximation. The variables $x_i$ are assumed i.i.d. Gaussian but limited to the value of the maxima $z$. This is equivalent to conditioning the Gaussian distribution on $|x| < z$.

$$p(x) = \frac{\frac{1}{\sqrt{2\pi}} e^{-\frac{x^2}{2}}}{\int_{-z}^{z} \frac{1}{\sqrt{2\pi}} e^{-\frac{x^2}{2}} dx}; \quad |x| < z, \quad p(x) = 0 \text{ otherwise}$$

So the variance of $x_i$ is the following function of $z$

$$Var(z) = \frac{\int_0^z x^2 e^{-\frac{x^2}{2}} dx}{\int_0^z e^{-\frac{x^2}{2}} dx} \tag{10}$$

We verified (10) by simulation of $N$-tuples of Gaussian variables for $N$=100 and found it accurate, see Figure 4.

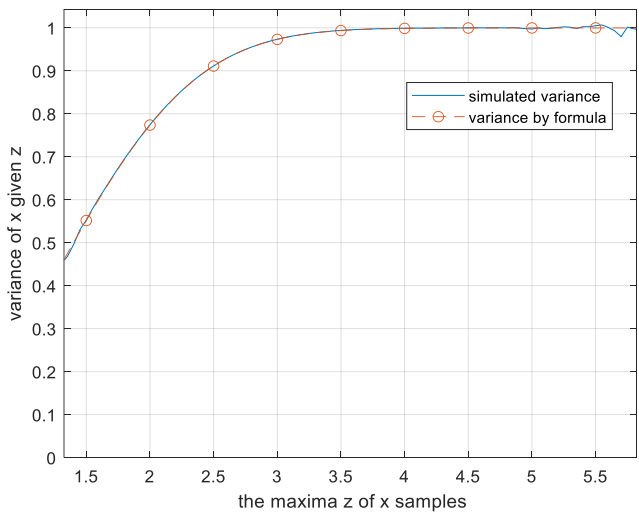

**Figure 4.** Variance of unordered $N$=100 Gaussian samples $x_i$, given their maxima $z$=max($|x_i|$) with the maximal sample excluded.

The length $L$ of the vector in the analysis below is not a random variable but rather the square root of the sum of the variances, for large $N$ this is a good approximation and dropping the randomness forms a lower bound on the entropy as explained below.

The length $L$ of the normalized vector is then:

$$L(z) = \sqrt{\frac{P \cdot Var(z) \cdot (N-1)}{z^2} + 1}$$

We compute the volume $V_x$ of the peak limited convex body with a unit power $P$=1, using ( 8) and the PDF in ( 9)

$$V_x = V_N^u \cdot E(L^N) = V_N^u \cdot \int_0^\infty p_z(z) L(z)^N dz \qquad (11)$$

This is a lower bound since $L(z)$ is assumed constant and only its root mean square is used. The accurate expression would replace in ( 11) the term $L(z)^N$ by $E[L(z)^N]$. This would increase $V_x$ by Jensen's inequality since $L(z)$ is positive and raised to a high power in the last equation. Thus, the lower bound evaluated at $P$=1 is

$$V_x = V_N^u \cdot \int_0^\infty p_z(z) \left(\frac{Var(z) \cdot (N-1)}{z^2} + 1\right)^{\frac{N}{2}} dz \qquad (12)$$

This is evaluated by numerical integration. The integrand of ( 12) indicates the range of the maxima $z$ most contributing to the capacity, in Figure 5 we plot the integrand, normalized by its maxima, for each $N$:

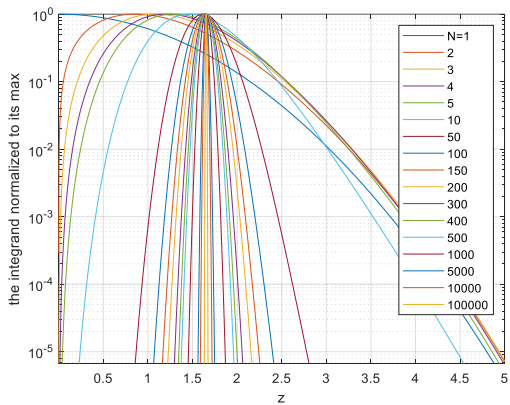

**Figure 5.** The integrand of equation ( 12) for different values of $N$.

So, for $N$>50, the system selects signals with relatively low peak values $z$ of about 1.7 and not increasing much with the length $N$ of the signal. This peak occurs, for $N$>100, with extremely low probability as seen in Figure 3. This implies that the signals contributing significantly to the integral ( 12) are rare. The power loss ratio relative to the APL system is obtained by inserting the result of ( 12) into ( 7). The power loss is shown in Figure 6.

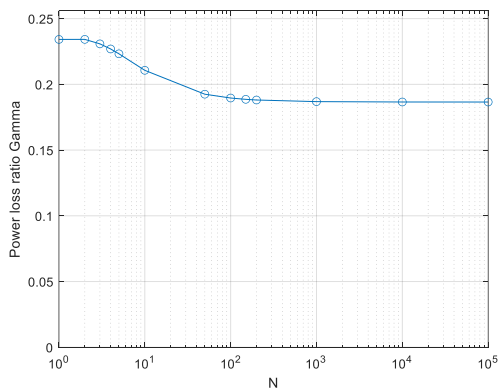

**Figure 6.** Computed lower bound on the power loss ratio $\gamma$, initial analysis limiting the power of Nyquist rate samples only.

The initial analysis here, which peak-limits only the Nyquist-rate samples, addresses the same discrete symbols problem as [4],[5] and[6] which present power loss of $\gamma$=0.2342 = 2/πe, see Appendix A. As explained above, the result in the last figure is a lower bound which explains the gap from our result to the correct 2/πe. Only the values for very low $N$ approach the exact value. Still, the result is near enough to the exact one to allow qualitative understanding of the numerical evaluation presented further below.

The volume-based evaluation of capacity developed here is tight at asymptotically high SNR as it is in the one-dimensional case [16] by the following. The capacity is

$$C = \max I(\boldsymbol{x}; \boldsymbol{y}) = h(\boldsymbol{y}) - h(\boldsymbol{n}) \tag{13}$$

Our lower bound ( 3),( 4) has a higher value then $h(\boldsymbol{x}) - h(\boldsymbol{n})$, see the derivation in Appendix A, so, to show that it is tight it suffices to show that at high SNR we have $h(\boldsymbol{x}) \cong h(\boldsymbol{y})$. The volume occupied by all vectors $\boldsymbol{x}$ is ( 11) which integrates the DFO of $\boldsymbol{x}$, denoted $L$, raised to the power of $N$, over all $\boldsymbol{x}$. For any $\boldsymbol{x}$, denoted $\boldsymbol{x}^i$, the DFO is $L \geq \sqrt{P}$. Now $\boldsymbol{y}$ cannot occupy a volume significantly larger than $V_x$ because the added noise increases the DFO in the direction of $\boldsymbol{x}^i$ only by a small multiple of $\sigma_n$ which is infinitely small relative to $\sqrt{P}$ at asymptotically large SNR. The volume $V_y$ occupied by all vectors $\boldsymbol{y}$ enforces $h(\boldsymbol{y}) \leq \log(V_y)$. Thus $h(\boldsymbol{x}) \cong h(\boldsymbol{y})$ holds and the lower bound is tight at high SNR.

## 3.3 Refinement to continuous signals

In this subsection we replace the peak power limit on the Nyquist rate samples $\boldsymbol{x}$ by peak power limit on the whole continuous signal $x(t)$. In Prasad [25] presents an authoritative and very convenient approximation to the PDF of PAPR of a <u>complex</u> signal in its (6.4). It is derived from (6.3), similar to our ( 9), by increasing $N$ by a factor of α=2.8, which is equivalent to replacing the infinity of correlated values in each Nyquist interval by α uncorrelated samples. This yields in our case of a <u>real-valued</u> signal the continuous version of CDF and PDF in ( 9) of $z_c$.

$$F_{Z_c}(z_c) = \left[F_{|X|}(z_c)\right]^{\alpha N} \tag{14}$$

And

$$p_{z_c}(z_c) = \alpha N[1 - 2Q(z_c)]^{\alpha N - 1} \frac{2}{\sqrt{2\pi}} e^{-\frac{z_c^2}{2}} \tag{15}$$

There are more advanced attempts to approximate $F_{Z_c}(z_c)$ such as [22], however those yielded still approximations and very involved expressions. The approximation ( 15) we use needs a verification by simulation. We found the following α a good approximation: α=2.3 for N=101, α=2.8 for N=1001, α=2.9 for N=10001 and N=100001. With N=1001, α=2.8 we get Figure 7.

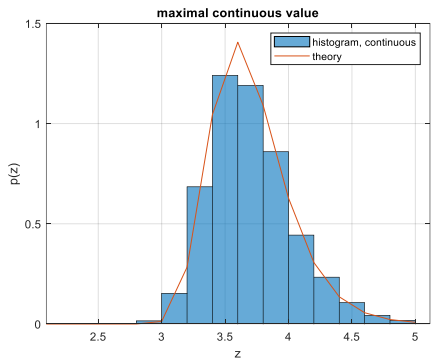

**Figure 7**. PDF of a maxima of a band limited Gaussian unit power random process over N=1001 Nyquist intervals.

Plugging ( 15) into ( 12) and ( 7) yields the power loss due to the peak limit in Figure 8.

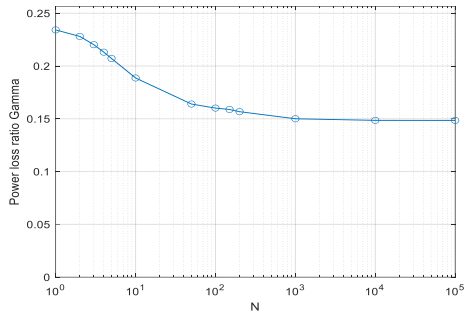

**Figure 8.** Computed lower bound on the power loss ratio γ, continuous real-valued PPL signalling.

**Conjecture 1:** *The power efficiency of 0.15 presented in Figure 8 is a lower bound on γ for low pass signals.*
**Explanation:** The only approximations used were ( 15) which is similar to [25] and verified numerically and the approximation (10) verified numerically in Figure 4. Both the approximations seem sound. The result of 0.15 is a lower bound on γ as explained below ( 11).

In the next section, to avoid all the approximations used for the analysis above, we shall evaluate the volume $V_x$, instead of the statistical expectation ( 11), by generating the vectors **x** at random and estimating $V_x$ by averaging via ( 8). We learn from the analysis the following lessons on the number of vectors required in the Monte Carlo evaluation. For N>100, the main contribution to $V_x$ and to capacity is from not too high values of $z$ which have very low probabilities as seen by combining Figure 5 with Figure 7. That is, the rare vectors the peak power of which is not too high contribute most to $V_x$. Generating at random enough vectors the probability of which is very low requires sufficient number of random vectors. As seen in Figure 3 and Figure 7, for N>100, the PDF $p(z)$ of a maxima of Gaussian process decreases sharply with decreasing $z$. So it is of interest to evaluate the power ratio γ with the volume in ( 12) integrated only over $z>z_{min}$ and plotting it as a function of $p(z<z_{min})$, such a plot will enable us to assess the number of simulated sequences required for convergence as follows: select the value of $z_{min}$ small enough to enable evaluating γ correctly using Figure 9, in which $p(z<z_{min})$ is presented rather than $z_{min}$, and use number of vectors larger than $1/p(z<z_{min})$ .

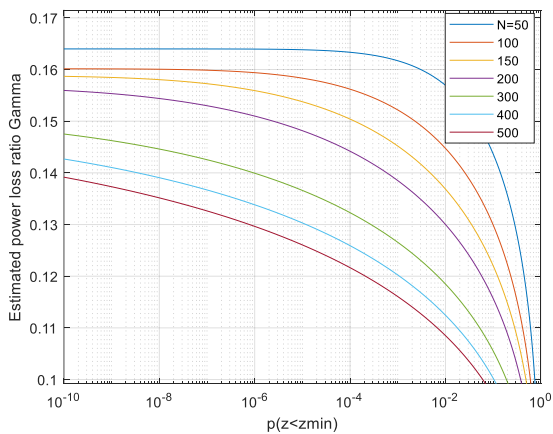

**Figure 9.** The ratio γ when vectors with a peak smaller then $z_{min}$ are discarded

Thus for *N*=50 about $10^4$ simulated sequences are required, for *N*=200 more than $10^8$ should be used. With $10^7$ simulated vectors we can expect reliable results up to *N* of about 100. To reach a reliable result we need to simulate enough vectors to be on the horizontal section of the curve. A useful criterion can be stable results under ten-fold change of p($z<z_{min}$). A practical method to examine this is discarding the 10 most contributing vectors and permitting only a small change in γ. When using the Monte Carlo evaluation with a too large *N*, the estimates will be lower than the true values because the low- probability vectors which contribute most to the capacity will be missed.

# 4 Monte Carlo evaluation

The Monte Carlo estimation avoids all the approximations used in the analysis section by generating the vectors **x** at random and estimating the expectation by averaging as explained above. The evaluation uses importance sampling as presented in Appendix B to accelerate convergence. The signals are evaluated as to be compatible with CP-FDE signaling, that is, generating $N$ Nyquist rate samples at random, oversampling while keeping the sequence duration at $NT$, performing FFT, brick-wall filtering in the frequency domain and IFFT. This is the classical transmit side processing of CP-FDE prior to adding the cyclic prefix. Each sequence is scaled as to have max($|x(t)|$)=1, the length $r$ of the scaled and sampled **x** is calculated and processed by the Monte Carlo equivalent of ( 8). As a verification, the discrete symbols case, that is, power limiting only the Nyquist rate samples and not the signal in between, [16] evaluates immediately to the correct $\gamma=2/\pi e$. To evaluate our continuous system oversampling ratio of 30 samples per Nyquist interval is used. We begin with sequence length of N=101 predicted to converge well by the analysis. The results are plotted in Figure 10.

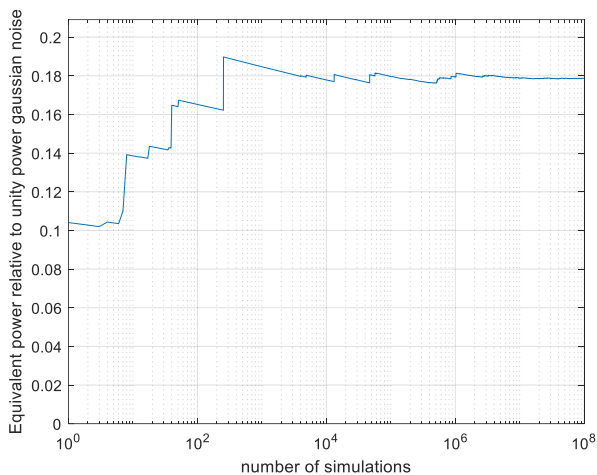

**Figure 10.** Monte Carlo evaluated power loss ratio $\gamma$, CP-FDE signalling, N=101, as a function of number of simulated vectors $N_{sim}$.

The result of $\gamma=0.18$ is somewhat larger than the lower bound of 0.15 computed by analysis in Figure 8, and there is a convergence after $N_{sim}= 10^6$ simulations. The convergence follows roughly the expectations based on Figure 9. For example, at $10^2$ simulations, $\gamma$ is about 0.02 below its final value as it is approximately for $p(z<zmin)=10^{-2}$ in Figure 9. In Figure 11 we present the results with the same simulation runs if the most contributing runs out of the total $10^8$ are discarded.

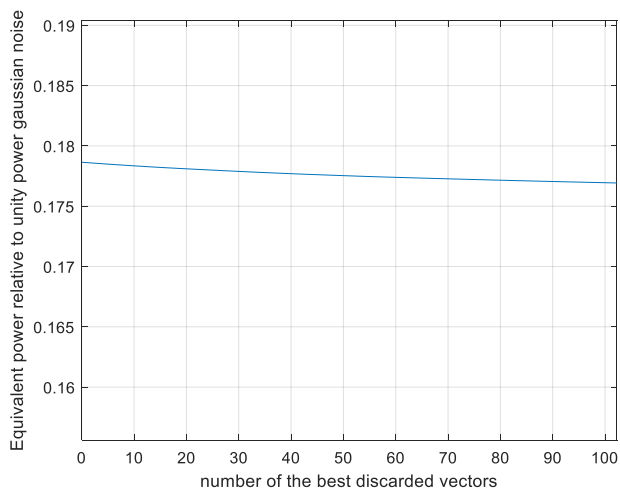

**Figure 11** Monte Carlo evaluated power loss ratio $\gamma$, N=101, as a function of number of the most contributing vectors discarded.

As seen, discarding the ten most contributing runs degrades the performance less than 0.2% and discarding 100 vectors degrades by 1% indicating a reliable convergence. This is comparable to rough expectation based on Figure 9. Next, we estimate $\gamma$ as a function of the sequence length $N$ using $N_{sim}=10^8$ simulated vectors per point. The results are plotted in Figure 12.

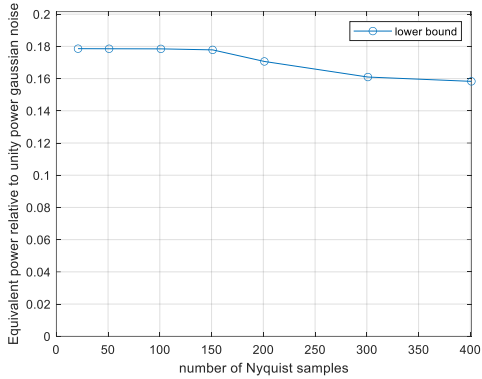

**Figure 12.** Monte Carlo evaluated power loss ratio $\gamma$, CP-FDE signalling, as a function of number of signal duration in Nyquist intervals $N$.

The $\gamma$ is above 0.177 for $N$ up to 101 and then starts to decrease. In Figure 13 we show the performance of discarding the $p$(discard) x $N_{sim}$ most contributing runs.

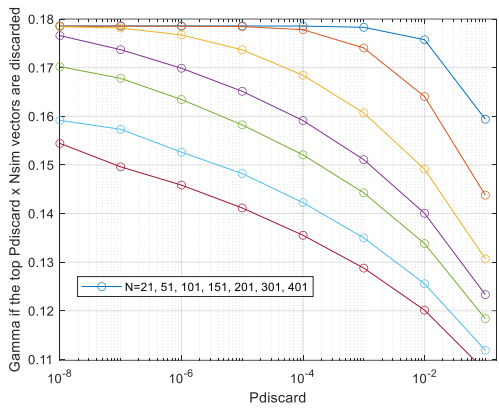

**Figure 13.** Monte Carlo evaluated power loss ratio $\gamma$, CP-FDE signalling, if the most contributing vectors the probability of which is *Pdiscard* are discarded.

The behaviour is qualitatively as predicted in Figure 9, that is, the decrease in the estimated results with rising $N$ is partly due to insufficient number of simulation runs, the most contributing vectors becoming rarer as $N$ rises. And, for lengths up to 101 Nyquist intervals the evaluation is well converged, that is, the results remain the same even if the 10 most contributing runs are discarded as seen in Figure 14.

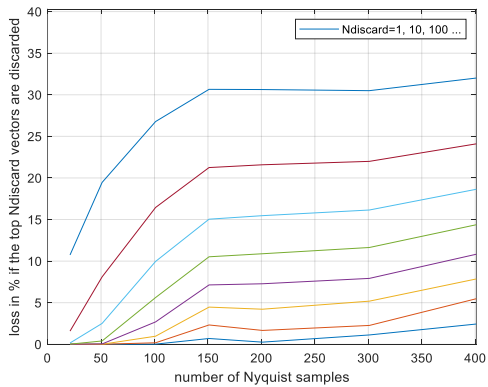

**Figure 14.** Monte Carlo evaluated power loss ratio γ, CP-FDE signalling, if the most contributing *Ndiscard* vectors are discarded.

# 5 Band pass signals

We extend the analysis to bandpass signals represented in the complex-valued baseband. We reuse the previous notation with modifications as follows. The encoder produces a complex-valued low-pass signal *x(t)* in the frequencies |*f*|<0.5*B*. The noise is complex-valued with power spectral density $N_0$ (two-sided). The signal is peak-limited, that is, $|x(t)| \leq \sqrt{P}$ for all $t$. The signal to noise ratio is defined as $\rho = \frac{P}{BN_0}$. The equations are updated as follows, see Appendix A. The classical capacity per Nyquist rate sample of the APL channel is

$$C_a = \log_2\left(\frac{P}{N_0 B} + 1\right) \tag{16}$$

The lower bound γ on the power loss ratio of the PPL channel relative to the APL channel remains

$$\gamma = \frac{P^e}{P} \leq 1 \tag{17}$$

where $P^e$ denotes the entropy power of **x** defined now as

$$P^e = \frac{1}{\pi e} \cdot e^{\frac{1}{N} h(x)} \tag{18}$$

It is shown in Appendix A that the following holds:

$$C \geq \log_2\left(\frac{\gamma \cdot P}{N_0 B} + 1\right) \tag{19}$$

In the PPL complex-valued channel the power ratio γ is shown in Appendix A to be

$$\gamma = \frac{V_x^{\frac{1}{N}}}{\pi e} \tag{20}$$

evaluated with *P*=1.

To compute the upper bound on capacity by power limiting only the Nyquist rate samples and not the signal in between, each complex sample of the entropy-maximizing distribution is uniformly distributed over a disk with radius of $\sqrt{P}$ yielding $V_x = \left(\pi\sqrt{P}^2\right)^N = (\pi P)^N$ and the ratio γ which is both an upper bound for our continuous signals case and an accurate value for the discrete power limited problem in [26].

$$\gamma = \frac{1}{e}, \tag{21}$$

this value falls indeed correctly between the Smith-based lower and upper bounds in Fig. 2 of [26], thus tightening the lower bound of [26] at high SNR.

Denote the maxima of $|x(t)|$ over $N$ Nyquist intervals as z and $w=z^2$. The work [25] reminds us that $w_s=|x(t)|^2$ is central chi-square distributed with two degrees of freedom (scaled to a unity mean) with $p_{w_s}(w_s) = e^{-w_s}$ and then provides a good approximation verified by simulations on the CDF of w:

$$F_w(w) = (1 - e^{-w})^{\alpha N} \tag{22}$$

with α=2.8. By differentiation

$$p_w(w) = \alpha N (1 - e^{-w})^{\alpha N - 1} e^{-w} \tag{23}$$

Equation (10) is replaced by:

$$E(w) = \frac{\int_0^w w_s e^{-w_s} dw_s}{\int_0^w e^{-w_s} dw_s} \tag{24}$$

where $E$ denotes expectation. The last equation was verified numerically by simulation of *N*-tuples of $w_s$ variables with N=100 and plotted in Figure 15.
.

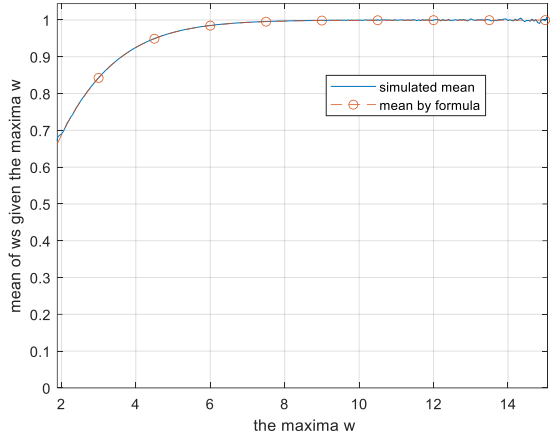

**Figure 15.** Mean of unordered $N$=100 samples $w_s$ given their maxima $w$ with the maximal sample excluded. Complex valued signals.

The typical length $L$ of the vector normalized to unity $\max(|x(t)|)$ will be then:

$$L(w) = \sqrt{\frac{E(w) \cdot (N-1)}{w} + 1} \qquad (25)$$

The volume $V_x$ is now $\quad V_{2N}^u \cdot E(L^{2N})$ because the signals are complex which doubles the dimension of the convex body. The volume $V_x$ of the peak limited convex body with a unit power $P$=1 using the PDF $p_w$ ( 23) is

$$V_x = V_{2N}^u \cdot \int_0^\infty p_w(w) L(w)^{2N} dw \qquad (26)$$

This is a lower bound as in the real-valued case. It is evaluated by numerical integration and the power loss ratio relative to the APL system is evaluated by ( 20). In the discrete symbols scenario, that is, power-limiting the Nyquist rate samples only, the result is as in Figure 16.

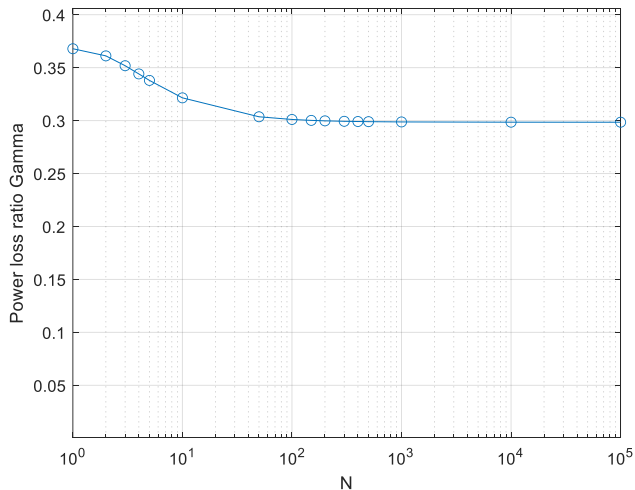

**Figure 16.** Computed lower bound on power loss ratio $\gamma$, initial analysis limiting the power of Nyquist rate samples only. Complex-valued signals.

As in the real-valued signals case, the result is a lower bound while the exact value is $\frac{1}{e} \cong 0.368$ , reached only at very low $N$. In the continuous case the lower bound on power efficiency $\gamma$ is plotted in Figure 17.

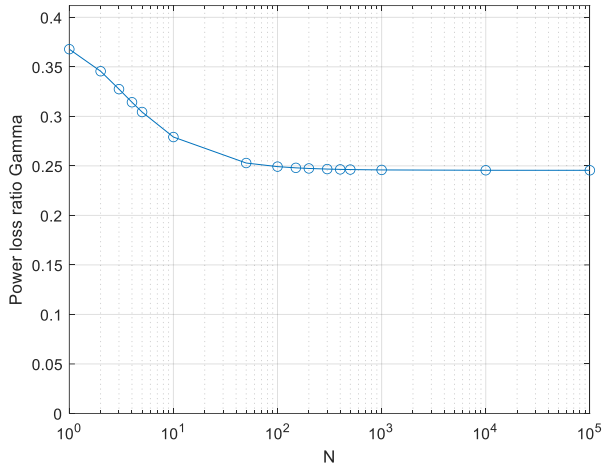

**Figure 17.** Computed lower bound on power loss ratio γ, continuous PPL signalling. Complex-valued signals.

An estimate of the number of vectors needed as a function of *N*, Figure 18, is generated by the same method as Figure 9 in the real signals case.

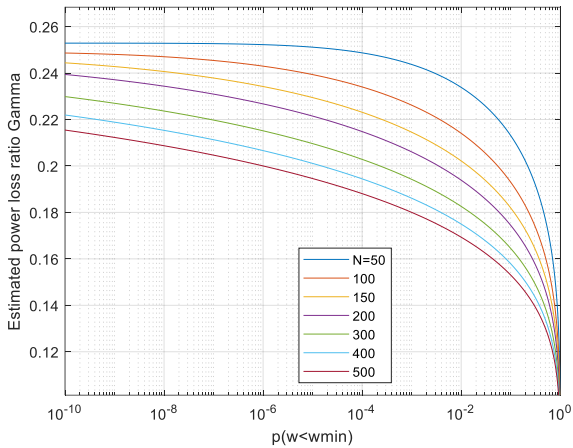

**Figure 18.** The ratio γ when vectors with a peak smaller then $w_{min}$ are discarded, complex-valued signals.

Comparing with Figure 9, it is evident that more vectors are required in the complex case, e.g. N=50 requires about $10^5$ vectors while N=101 requires about $10^8$.

**Conjecture 2:** *The power efficiency of 0.245 presented in Figure 17 is a lower bound on γ for band-pass signals.*
**Explanation:** the only approximations used were ( 22) adopted from [25] and the truncation in ( 24) which was verified numerically. Both the approximations seem sound. The result of 0.245 is a lower bound on γ as explained below ( 11).

The Monte Carlo estimation, as in the real-valued case, avoids all the approximations used in the analysis. The signals are evaluated as to be compatible with CP-FDE signaling. As a verification, the discrete symbols case evaluates immediately to the correct γ=1/e. To evaluate our continuous system, an oversampling ratio of 30 samples per Nyquist interval is used. We begin with sequence lengths of 51*T* and 101*T* predicted to converge by the analysis. The results are shown in Figure 19.

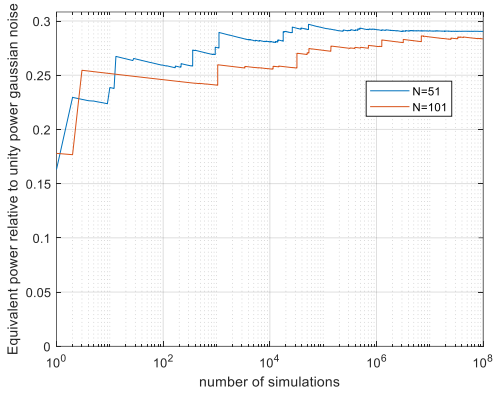

**Figure 19.** Monte Carlo evaluated power loss ratio γ, CP-FDE signalling, *N*=51 and 101 as a function of number of simulated vectors. Complex-valued signals.

The result N=101, γ=0.285 is slightly larger than the computed lower bound of 0.245 in Figure 17, and there is a convergence after $10^6$ and $10^7$ simulations for N=51 and N=101 respectively. In Figure 20 we present the results with the same simulation runs if the most contributing runs out of the total $10^8$ are discarded.

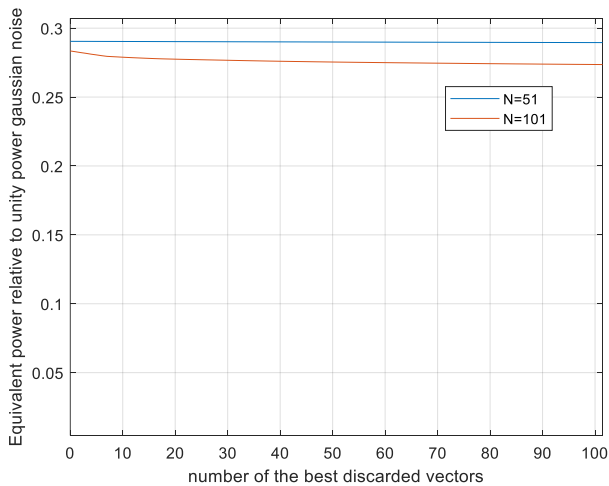

**Figure 20.** Monte Carlo evaluated power loss ratio γ, N=101 as a function of number of the most contributing vectors discarded. Complex-valued signals.

As seen, for N=101, discarding the ten most contributing runs degrades the performance by less than 2%. The results for N=51 are stable even if 100 most contributing runs are discarded. This behaviour is comparable to rough expectation based on Figure 18. Next, we estimate γ as a function of the sequence length *N* using $N_{sim}$=$10^8$ vectors per point. The results are presented in Figure 21.

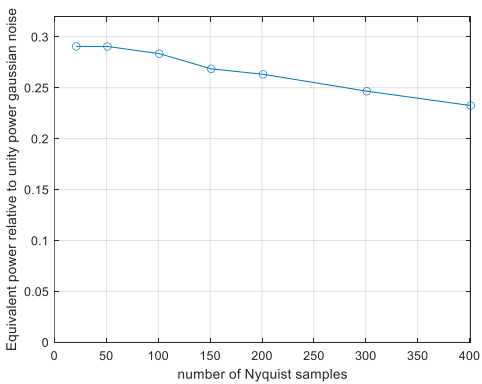

**Figure 21.** Monte Carlo evaluated power loss ratio γ, CP-FDE signalling, as a function of number of signal duration in Nyquist intervals *N*. Complex-valued signals.

In Figure 22 we show the performance when discarding the $P(\text{discard}) \times Nsim$ most contributing runs:

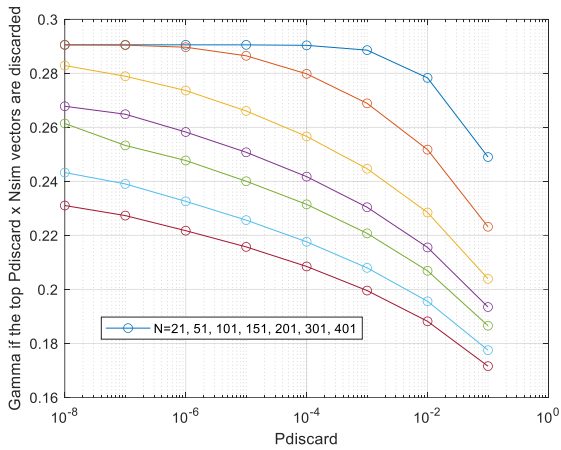

**Figure 22.** Monte Carlo evaluated power loss ratio γ, CP-FDE signalling, if the most contributing vectors the probability of which is *Pdiscard* are discarded. Complex-valued signals.

The behaviour is qualitatively as predicted in Figure 18, that is, the decrease in the estimated results is in part due to not enough simulation runs, the most contributing vectors becoming rarer as *N* rises. The convergence of the Monte Carlo evaluation can be demonstrated also by Figure 23.

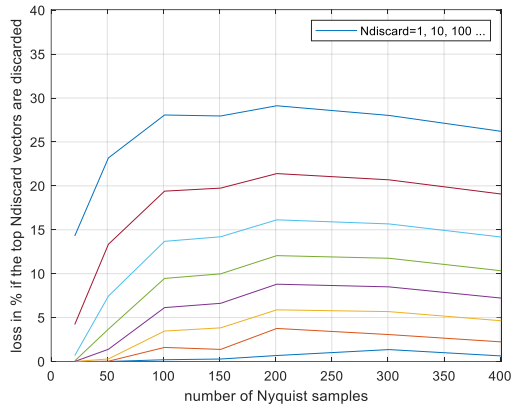

**Figure 23.** Monte Carlo evaluated power loss ratio γ, CP-FDE signalling, if the most contributing vectors the probability of which is Pdiscard are discarded. Complex-valued signals.

The *N*=51 result γ=0.29 is reliable, with $N_{sim}$=10⁸ corresponding to *Pdiscard* of 10⁻⁸, it is on the horizontal part of the curves in Figure 18 and in Figure 22 and is stable if 100 most contributing vectors are discarded. The result at N=101 is probably slightly lower than the true γ, it losses 1.6% if 10 most contributing vectors are discarded. The result γ=0.29 is the γ of the CP-FDE signaling at signal duration of 51*T* and the values in Figure 21 are lower bounds applicable to the CP-FDE signaling.

# 6 Conclusions

The important problem of the capacity of the PPL channel was investigated. We focussed on the power efficiency γ which provides a lower bound on capacity which is tight at asymptotically high SNR. The results are summarized in Table 1. We showed that the lower bounds on γ are about 3.35 and 8.6 times higher than previously known in the low-pass and in the band pass cases respectively. The gap to the upper bounds is narrowed to less than 2 dB. The numerical results on γ are valid for the practical CP-FDE signaling with limited transmission duration while the general analytical results are lower bounds on γ as explained below ( 11) and rely on two approximations rendering them a conjecture. The lower bounds based on γ via ( 3) are tight at high SNR and show that the peak power constraint causes, at high SNR, power loss of about 6 dB in the bandpass case and a little more in the low pass case.

**Table 1**, bounds on power efficiency γ of peak limited signals

| The bound | results of [5] | results of [6] | This work, general result, but remains a conjecture | This work | This work, CP-FDE signaling, Duration of 101 Nyquist intervals |
|---|---|---|---|---|---|
| **Low pass lower bound** | $0.0361 = \pi/32e$ | 0.04470 | $\gamma > 0.15$ | | $\gamma = 0.18$ |
| **Low pass upper bound** | $0.2342 = 2/\pi e$ Presented also in [4]. | $0.2342 = 2/\pi e$ | | | |
| **Band pass lower bound** | $\frac{\pi^2}{128e} = 0.0284$ | | $\gamma > 0.245$ | | $\gamma = 0.285$ |
| **Band pass upper bound** | | | | $\frac{1}{e} \cong 0.368$ | |

Future work: The above stated upper bounds are based on peak-power-limiting of only the Nyquist rate samples which are assumed to be independent. It would be interesting to account also for the correlation, which is implied on these samples, as is reflected in [15], as to sharpen the upper bounds. This might be addressed via a Multiple Input Multiple Output (MIMO) structure, where both constraints should be addressed. The MIMO setting might also be relevant for a super Nyquist sampling, again accounting for the fact that all samples are peak-power limited, see [15], [27] for relevant results that might be used.

Also, our analysis showed that the capacity achieving signals are a small proportion of Gaussian bandlimited signals and this proportion vanishes with growing signal duration. The same holds on codewords of error correcting codes, the codewords are a vanishing proportion of all random binary sequences. The peak limit selects the appropriate signals resembling the effect of the parity check matrix of a binary error correcting code. Thus, further work should seek structures of PPL signals approaching the channel capacity similar to the vast work done in recent decades on error correcting codes.

## Appendix A - bounds

We summarize the bounding technique used e.g. in [10] for completeness. We shall start with real-valued signals. Denote differential entropy as $h$. The channel is an AWGN channel with scalar-valued real symbols ( 1 ). The noise has power of $\sigma_n^2 = N_0 B$ and per-sample entropy of $h(n) = 0.5 \log(\sigma_n^2 \cdot 2\pi e)$. The capacity per sample is

$$C = \frac{1}{N}[h(\boldsymbol{y}) - h(\boldsymbol{y}|\boldsymbol{x})] = \frac{1}{N}[h(\boldsymbol{y}) - h(\boldsymbol{n})]$$

where the bold symbols denote vectors and $N$ is the number of Nyquist rate samples. By EPI:

$$e^{\frac{2}{N}h(\boldsymbol{y})} \geq e^{\frac{2}{N}h(\boldsymbol{x})} + e^{\frac{2}{N}h(\boldsymbol{n})} \qquad (27)$$

$$h(\boldsymbol{y}) \geq \frac{N}{2}\log\left(e^{\frac{2}{N}h(\boldsymbol{x})} + e^{\frac{2}{N}h(\boldsymbol{n})}\right)$$

and

$$e^{\frac{2}{N}h(\boldsymbol{n})} = \sigma_n^2 \cdot 2\pi e$$

So the per sample capacity C is

$$C \geq \frac{1}{2}\log\left(e^{\frac{2}{N}h(\boldsymbol{x})} + \sigma_n^2 \cdot 2\pi e\right) - \frac{1}{2}\log(\sigma_n^2 \cdot 2\pi e)$$

$$C \geq \frac{1}{2} \log \left( \frac{\frac{1}{2\pi e} \cdot e^{\frac{2}{N}h(x)}}{\sigma_n^2} + 1 \right) \qquad (28)$$

We use the entropy power of $x$ definition $P^e = \frac{1}{2\pi e} \cdot e^{\frac{2}{N}h(x)}$, yielding, in bits per sample,

$$C \geq \frac{1}{2} \log_2 \left( \frac{P^e}{\sigma_n^2} + 1 \right) \qquad (29)$$

followed by the pre-SNR factor ( 3),

( 4). The ratio $\gamma$ in ( 4) is the lower bound on the power loss ratio of the PPL channel relative to the APL channel and is valid at all SNRs. It is used in [6] as the pre-SNR factor. Now $h(x)=\log(V_x)$ where $V_x$ is the volume of the convex body if the distribution of $x$ in is the entropy-maximizing uniform one. This yields ( 6). It follows from the peak power limit $P$ that $V_x$ is proportional to $P^{\frac{N}{2}}$. Then, by ( 6), $P^e$ is linearly proportional to $P$ and volume-derived $\gamma$ is invariant with respect to $P$. Therefore, $\gamma$ can be evaluated with $P=1$ with no loss of generality.

To provide an upper bound on $\gamma$, the peak power limit can be applied on the Nyquist rate samples only and not on the signal in between as done in [5]. In this case the signal samples are limited to $-\sqrt{P} < x < \sqrt{P}$, the convex body is an $N$-cube with volume $(2\sqrt{P})^N$ and ( 7) yields the power loss $\gamma$ of $\frac{2}{\pi e}$ presented in [5].

**Extension to complex-valued signals:**

The symbols are complex. The real and imaginary components of the noise have power of $\sigma_n^2 /2 = N_0B/2$ each and per-symbol noise entropy is $h(n) = \log(\sigma_n^2 \cdot \pi e)$. The per-symbol capacity is

$$C = \frac{1}{N} [h(y) - h(n)])$$

The EPI formula as in ( 27) holds for $N$-dimensional real vectors, for $N$-dimensional complex vectors it is:

$$e^{\frac{1}{N}h(y)} \geq e^{\frac{1}{N}h(x)} + e^{\frac{1}{N}h(n)}$$

yielding

$$C \geq \log \left( \frac{\frac{1}{\pi e} \cdot e^{\frac{1}{N}h(x)}}{\sigma_n^2} + 1 \right) \qquad (30)$$

For a gaussian signal with power $P$ this turns to the familiar ( 16).
We can denote the entropy power $P^e$ of the complex-valued x as in (18) and define $\gamma$ by ( 4) yielding ( 19).
As in the real valued case, the maximal entropy is $h(x) = \log(V_x)$ where $V_x$ is the volume of the convex body. This yields, for the uniform entropy-maximizing distribution of x:

$$P^e = \frac{1}{\pi e} V_x^{\frac{1}{N}} \qquad (31)$$

and

$$C \geq \log \left( \frac{\frac{1}{\pi e}V_x^{\frac{1}{N}}}{\sigma_n^2} + 1 \right) \qquad (32)$$

and ( 20) follows.

## Appendix B. Importance sampling in the n-cube

As explained, generating signals at random on the $N$-sphere yields Gaussian-distributed samples. This enabled analytical results and straightforward simulation, however there is a problem. As explained above, with no oversampling, the optimal signal is distributed uniformly in the hypercube, the signal volume is concentrated near the $N$-cube corners and the corner regions are visited by the random signals, which are uniform on the on the $N$-sphere, very rarely. Thus, the evaluation by simulation converges very slowly. We shall accelerate the estimation by importance sampling, generating the random signals uniformly in the hypercube instead on the unit-radius hypersphere. This accelerates the evaluation to some degree even if the signal region in the continuous case is definitely not a hypercube. The principles are:

The signals are generated uniformly distributed in the volume of an $N$-cube by randomly generating each sample $x_i$ uniformly in the [-1 , 1] interval. For each signal instance a probability factor $Pc$ is computed, equal to probability of this particular angle if the signals were generated uniformly over angles, that is uniformly over the unity radius $N$-sphere, divided by the probability if the signals were generated uniformly in the $N$-cube. $P_c$ is the ratio of the volume of the hyperball enclosed in an infinitesimal angle sector around the selected signal to the volume of the hypercube enclosed in the same-width angle sector corrected by the volume ratio of the hypercube $2^N$ and the hypersphere $V_N^u$. Thus, denoting by $L_c$ the DFO of the hypercube surface in the chosen direction,

$$P_c = \frac{1}{L_c^N} \frac{2^N}{V_N^u}$$

The volume evaluation ( $8$ ) is modified according to the importance sampling method to:

$$V_Z = \mathrm{E}_\theta(V_N^u \cdot r^N \cdot P_c) = \mathrm{E}_\theta\left(V_N^u \cdot r^N \cdot \frac{1}{L_c^N} \frac{2^N}{V_N^u}\right) = \mathrm{E}_\theta\left\{\left(\frac{2 \cdot r}{L_c}\right)^N\right\} \qquad (33)$$

where the expectation is carried with signals distributed uniformly over the N-cube. Intuitive check: suppose the signal body is an $N$-cube with samples $-1 < x_i < 1$. Then always $r = L_c$ since $r$ is scaled by the peak limiting to touch the hypercube surface and the volume will be $2^N$ as it should be.

**Extension to the complex-valued case:**

The signals are now generated uniformly distributed in the volume of an $2N$-dimensional body by randomly generating each of the $N$ complex samples $x_i$ independently and uniformly over area in the complex plane enclosed by a circle with a unity radius reflecting $| x_i | < 1$ and area of $\pi$. The last two equations are updated accordingly accounting for the volume of the $2N$-dimensional body of $\pi^N$ and for the number of dimensions being $2N$ since each $x_i$ is complex. $L_c$ is the DFO of the surface of the $2N$-dimensional body in the chosen direction.

$$P_c = \frac{1}{L_c^{2N}} \frac{\pi^N}{V_{2N}^u}$$

The volume evaluation ( $8$ ) is modified according to the importance sampling method to:

$$V_Z = \mathrm{E}_\theta(V_{2N}^u \cdot r^{2N} \cdot P_c) = \mathrm{E}_\theta\left\{\pi^N \left(\frac{r}{L_c}\right)^{2N}\right\} \qquad (34)$$

--------------------------------------------------------------------------------------------------------------------------------

**Acknowledgments:** This work was supported by the German Research Foundation (DFG) via the German-Israeli Project Cooperation (DIP), under Project SH 1937/1-1.